\documentclass{article}%
\usepackage{amsmath}
\usepackage{amsfonts}
\usepackage{amssymb}
\usepackage{graphicx}%
\begin{document}

\title{Entropic Priors\thanks{Presented at MaxEnt 2003, the 23th International
Workshop on Bayesian Inference and Maximum Entropy Methods (August 3-8,
Jackson Hole, WY, USA).}}
\author{Ariel Caticha$^{*}$ and Roland Preuss$^{\dagger}$\\$^{\ast}${\small Department of Physics, University at Albany-SUNY, }\\{\small Albany, NY 12222, USA.}\\$^{\dagger}${\small Center for Interdisciplinary Plasma Science, }\\{\small Max-Planck-Institut f\"{u}r Plasmaphysik, EURATOM Association, }\\{\small Boltzmannstrasse 2, D-85748 Garching bei M\"{u}nchen, Germany}}
\date{}
\maketitle

\begin{abstract}
The method of Maximum (relative) Entropy (ME) is used to translate the
information contained in the known form of the likelihood into a prior
distribution for Bayesian inference. The argument is guided by intuition
gained from the successful use of ME methods in statistical mechanics. For
experiments that cannot be repeated the resulting \textquotedblleft entropic
prior\textquotedblright\ is formally identical with the Einstein fluctuation
formula. For repeatable experiments, however, the expected value of the
entropy of the likelihood turns out to be relevant information that must be
included in the analysis. As an example the entropic prior for a Gaussian
likelihood is calculated.

\end{abstract}

\section{Introduction}

Among the methods used to update from a prior probability distribution to a
posterior distribution when new information becomes available there are two
that can claim the distinction of being systematic, objective, and of wide
applicability:\ one is based on Bayes' theorem (for applications to physics
see \cite{Dose03}) and the other is based on the maximization of (relative)
entropy \cite{Caticha03}. The choice between the two methods is dictated by
the nature of the information being processed.

Bayes' theorem should be used when we want to update our beliefs about the
values of quantities $\theta$ on the basis of observed values of data $y$ and
of the known relation between them -- the likelihood $p(y|\theta)$. The
posterior distribution is $p(\theta|y)\propto\pi(\theta)p(y|\theta)$. The
previous knowledge about $\theta$ is codified both in the prior distribution
$\pi(\theta)$ and also in the likelihood\ $p(y|\theta)$.

The selection of the prior is a difficult problem \cite{Kass96} because it is
not always clear how to translate our previous beliefs about $\theta$ into a
distribution $\pi(\theta)$ in an objective way. One approach that seems to
work, at least sometimes, is to rely on experience and physical intuition but
this becomes unreliable in situations of increasing complexity. Attempts to
achieve objectivity include arguments invoking symmetry -- generalized forms
of the principle of insufficient reason -- and arguments that seek to identify
that state of knowledge that reflects complete ignorance. The latter suggest
connections with the notion of entropy \cite{Jaynes68} and have led to
proposals for \textquotedblleft entropic priors\textquotedblright%
\ \cite{Skilling89, Rodriguez89}. This brings us to the second method of
processing information, the method of maximum entropy, which is designed for
processing information given in the form of constraints on the family of
posterior distributions \cite{Caticha03}.

In this paper we use entropic arguments to translate information into a prior
distribution \cite{Caticha03b}. Rather than seeking a totally non-informative
prior, we translate information that we do in fact have: the knowledge of the
likelihood function, $p(y|\theta)$, already constitutes valuable prior
information. The prior thus obtained is an \textquotedblleft entropic
prior.\textquotedblright\ The \emph{bare} entropic priors discussed here apply
to a situation where all we know about the quantities $\theta$ is that they
appear as parameters in the likelihood $p(y|\theta)$. It is straightforward,
however, to extend the method and incorporate additional relevant information
beyond that contained in the likelihood.

The first proposal of priors of this form is due to Skilling \cite{Skilling89}
for the case of discrete distributions. The second proposal, due to
Rodr\'{\i}guez \cite{Rodriguez89}, provided the generalization to the
continuous case and further elaborations \cite{Rodriguez98, Rodriguez02}. In
section 2 we give a derivation that is closer in spirit to applications of
ME\ to statistical mechanics. A difficulty with the case of experiments that
can be indefinitely repeated, which had been identified in \cite{Caticha00},
is diagnosed and resolved with the introduction of a hyper-parameter $\alpha$
in section 3. The analogy to statistical mechanics is important: the
interpretation of $\alpha$ as a Lagrange multiplier affects how $\alpha$
should be estimated and is an important difference between the entropic prior
proposed here and those of Skilling and Rodr\'{\i}guez. The example of a
Gaussian likelihood is given in section 4. In section 5 we collect our
conclusions and some final comments.

\section{The basic idea}

We use the ME method \cite{Caticha03} to derive a prior $\pi(\theta)$ for use
in Bayes' theorem $p(\theta|y)\propto p(y,\theta)=\pi(\theta)p(y|\theta)$. As
discussed in \cite{Caticha00}, since Bayes' theorem follows from the product
rule we must focus our attention on $p(y,\theta)$ rather than $\pi(\theta)$.
Thus, the relevant universe of discourse is the product $\Theta\times Y$ of
$\Theta$, the space of all $\theta$s, and the data space $Y$. This important
point was first made by Rodr\'{\i}guez \cite{Rodriguez98} but both our
derivation and final results differ from his \cite{Rodriguez98, Rodriguez02}.

To rank distributions on the space $\Theta\times Y$ we must first decide on a
prior $m(y,\theta)$. When nothing is known about the variables $\theta$ -- in
particular, no relation between $y$ and $\theta$ is yet known -- the prior
must be a product $m(y)\mu(\theta)$ of the separate priors in the spaces $Y$
and $\Theta$ because maximizing the relative entropy
\begin{equation}
\sigma\lbrack p]=-\int dy\,d\theta\,p(y,\theta)\,\log\frac{p(y,\theta
)}{m(y)\mu(\theta)},
\end{equation}
yields $p(y,\theta)\propto m(y)\mu(\theta)$.This distribution reflects our
state of ignorance: the data about $y$ tells us absolutely nothing about
$\theta$.

In what follows we assume that $m(y)$ is known because it is an important part
of understanding what data it is that has been collected. Furthermore, if the
$\theta$s are parameters labeling some distributions $p(y|\theta)$, then for
each particular choice of the functional form of $p(y|\theta)$ there is a
natural distance in the space $\Theta$ given by the Fisher-Rao metric
$d\ell^{2}=g_{ij}d\theta^{i}d\theta^{j}$, \cite{Amari85}
\begin{equation}
g_{ij}=\int dy\,p(y|\theta)\frac{\partial\log p(y|\theta)}{\partial\theta^{i}%
}\frac{\partial\log p(y|\theta)}{\partial\theta^{j}}. \label{Fisher metric}%
\end{equation}
Therefore the prior on $\theta$ is $\mu(\theta)=g^{1/2}(\theta)$ where
$g(\theta)$ is the determinant of $g_{ij}$.

Next we incorporate the crucial piece of information: of all joint
distributions $p(y,\theta)=\pi(\theta)p(y|\theta)$ we consider the subset
where the likelihood $p(y|\theta)$ has a fixed, known functional form. Notice
that this is an unusual constraint; it is not an expectation value. Note also
that the only information we are using about the quantities $\theta$ is that
they appear as parameters in the known likelihood $p(y|\theta)$, \emph{nothing
else}. But, of course, should additional relevant information (\emph{i.e.}, an
additional constraint) be known it should also be taken into account.

The preferred distribution $p(y,\theta)$ is chosen by varying $\pi(\theta)$ to
maximize
\begin{equation}
\sigma\lbrack\pi]=-\int dy\,d\theta\,\pi(\theta)p(y|\theta)\,\log\frac
{\pi(\theta)p(y|\theta)}{g^{1/2}(\theta)m(y)}~. \label{sigma[pi]}%
\end{equation}
$\,$ Assuming that both $\,\pi(\theta)$ and $p(y|\theta)$ are normalized the
result is
\begin{equation}
\pi(\theta)d\theta=\frac{1}{\zeta}\,\,e^{S(\theta)}g^{1/2}(\theta)d\theta
\quad\text{where}\quad\zeta=\int d\theta\,g^{1/2}(\theta)\,e^{S(\theta)},
\label{main}%
\end{equation}
and $S(\theta)$ is the entropy of the likelihood,
\begin{equation}
S(\theta)=-\int\,dy\,p(y|\theta)\log\frac{p(y|\theta)}{m(y)}. \label{Stheta}%
\end{equation}
The entropic prior eq.(\ref{main}) is our first important result: it gives the
probability that the value of $\theta$ should lie within the small volume
$g^{1/2}(\theta)d\theta$. The preferred value of $\theta$ is that which
maximizes the entropy $S(\theta)$ because this maximizes the scalar
probability density $\exp S(\theta)$. Note that eq.(\ref{main}) manifestly
invariant under changes of the coordinates $\theta$.

To summarize: for the special case of a fixed data space $Y$, that is, for
experiments that cannot be repeated, we have succeeded in translating the
information contained in the model -- the space $Y$, its measure $m(y)$, and
the conditional distribution $p(y|\theta)$ -- into a prior $\pi(\theta)$.

But for experiments that can be repeated indefinitely the prior (\ref{main})
yields nonsense and we have a problem. Indeed, let us assume that $\theta$ is
not a \textquotedblleft random\textquotedblright\ variable, its value is fixed
but unknown. For $N$ independent repetitions of an experiment, the joint
distribution in the space $\Theta\times Y^{N}$ is
\begin{equation}
p(y^{(N)},\theta)=\pi^{(N)}(\theta)\,p(y^{(N)}|\theta)=\pi^{(N)}%
(\theta)p(y_{1}|\theta)\ldots p(y_{N}|\theta),
\end{equation}
and maximization of the appropriate $\sigma^{(N)}$ entropy gives
\cite{Caticha00}
\begin{equation}
\pi^{(N)}(\theta)=\frac{1}{Z^{(N)}}\,g^{1/2}(\theta)\,e^{NS(\theta)},
\label{pi(n)}%
\end{equation}
which is clearly wrong. The dependence of $\pi^{(N)}$ on the amount $N$ of
data would lead us to a perpetual revision of the prior as more data is
collected. For large $N$ the data becomes irrelevant.

The problem, as we will see next, is not a failure of the ME method but a
failure to include all the relevant information. Indeed, when an experiment
can be repeated we actually know more than just $p(y^{(N)}|\theta
)=\,p(y_{1}|\theta)\ldots p(y_{N}|\theta)$. We also know that discarding the
values of say $y_{2},\ldots y_{N}$, yields an experiment that is
indistinguishable from the single, $N=1$, experiment. This \emph{additional}
information, which is expressed by $\int dy_{2}\ldots dy_{N}\,p(y^{(N)}%
,\theta)=p(y_{1},\theta)$ leads to $\pi^{(N)}(\theta)=\pi^{(1)}(\theta)$ for
all $N$. Next we identify a constraint that codifies this information within
each space $\Theta\times Y^{N}$.

\section{More information: the Lagrange multiplier $\alpha$}

For large $N$ the prior $\pi^{(N)}(\theta)$ in eq.(\ref{pi(n)}) reflects an
overwhelming preference for the value of $\theta$ that maximizes the entropy
$S(\theta)$. Indeed, as $N\rightarrow\infty$ we have
\begin{equation}
\langle S\rangle=\int d\theta\,\pi^{(N)}(\theta)S(\theta)\overset
{N\rightarrow\infty}{\longrightarrow}S(\theta_{\max})~,
\end{equation}
which is manifestly incorrect. This suggests that information about the actual
numerical value $\bar{S}$ of the expected entropy $\langle S\rangle$ is very
relevant (because if $\bar{S}$ were known the problem above would not arise)
and that we should maximize $\sigma^{(N)}$ subject to an additional constraint
on $\bar{S}$. Naturally, additional steps will be needed to estimate the
unknown $\bar{S}$. A similar argument justifying the introduction of
constraints in statistical physics is explored in \cite{Caticha03}.

We maximize the entropy%

\begin{equation}
\sigma^{(N)}[\pi]=-\int\,d\theta\,dy^{(N)}\,\pi(\theta)p(y^{(N)}|\theta
)\log\frac{\pi(\theta)p(y^{(N)}|\theta)}{g^{1/2}(\theta)\,m(y^{(N)})}%
\end{equation}
subject to constraints on $\langle S\rangle$ and that $\pi$ be normalized. (An
unimportant factor of $N^{d/2}$ has been dropped from the Fisher-Rao measure
$g^{(N)}(\theta)$.) The result is
\begin{equation}
\pi(\theta)=\frac{1}{\zeta}g^{1/2}(\theta)\exp\left[  (N+\lambda_{N}%
)S(\theta)\right]  ~.
\end{equation}
The undesired dependence on $N$ is eliminated if the Lagrange multipliers
$\lambda_{N}$ in each space $\Theta\times Y^{N}$ are chosen so that
$N+\lambda_{N}=\alpha$ is a constant independent of $N$. The resulting
entropic prior,
\begin{equation}
\pi(\theta|\alpha)=\frac{1}{\zeta(\alpha)}g^{1/2}(\theta)e^{\alpha S(\theta)}~
\label{main 2}%
\end{equation}
is our second important result. The prior $\pi(\theta|\alpha)$ incorporates
information contained in the likelihood plus information about
\begin{equation}
\langle S\rangle=\bar{S}(\alpha)=\frac{d}{d\alpha}\log\zeta(\alpha
)\quad\text{where}\quad~\zeta(\alpha)=\int d\theta~g^{1/2}(\theta)e^{\alpha
S(\theta)}~. \label{Sbar}%
\end{equation}

The last step would be to estimate $\alpha$ and $\theta$ from Bayes' theorem
\begin{equation}
p(\alpha,\theta|y^{(N)})=\pi(\alpha)\pi(\theta|\alpha)\frac{p(y^{(N)}|\theta
)}{p(y^{(N)})}\,,
\end{equation}
where $\pi(\alpha)$ is a prior for $\alpha$. However, if we are only
interested in $\theta$, we can just marginalize over $\alpha$ to get
\begin{equation}
p(\theta|y^{(N)})=\int d\alpha\,p(\alpha,\theta|y^{(N)})=\bar{\pi}%
(\theta)\frac{p(y^{(N)}|\theta)}{p(y^{(N)})} \label{entropic Bayes}%
\end{equation}
where
\begin{equation}
\bar{\pi}(\theta)=\int d\alpha\,\pi(\alpha)\pi(\theta|\alpha)~. \label{main 3}%
\end{equation}
The averaged $\bar{\pi}(\theta)$ is our final expression for the entropic
prior. It is independent of the actual data $y^{(N)}$ as it should.

Next we assign an entropic prior to $\alpha$. We start by pointing out that
$\alpha$ is not on the same footing and should not be treated like the other
parameters $\theta$ because the relation between $\alpha$ and the data $y$ is
indirect: $\alpha$ is related to $\theta$ through $\pi(\theta|\alpha)$, and
$\theta$ is related to $y$ through $p(y|\theta).$ Once $\theta$ is given, the
data $y$ contains no further information about $\alpha$. Since the whole
significance of $\alpha$ is derived purely from $\pi(\theta|\alpha)$,
eq.(\ref{main 2}), the relevant universe of discourse is $A\times\Theta$ with
$\alpha\in A$ and not $A\times\Theta\times Y^{N}$ as in \cite{Rodriguez89}
which requires the introduction of an endless chain of hyper-parameters.

We therefore consider the joint distribution $\pi(\alpha,\theta)=\pi
(\alpha)\pi(\theta|\alpha)$ and obtain $\pi(\alpha)$ by maximizing the
entropy
\begin{equation}
\Sigma\lbrack\pi]=-\int d\alpha\,d\theta\,\,\pi(\alpha,\theta)\log\frac
{\pi(\alpha,\theta)}{\gamma^{1/2}(\alpha)\,g^{1/2}(\theta)}\,
\label{Sigma prime}%
\end{equation}
where $\gamma^{1/2}(\alpha)$ is determined below. Since no reference is made
to repeatable experiments in $Y^{N}$ there is no need for any further
constraints -- and no further hyper-parameters -- except for normalization.
The result is
\begin{equation}
\pi(\alpha)=\frac{1}{z}\gamma^{1/2}(\alpha)e^{s(\alpha)}\,, \label{main 4}%
\end{equation}
where using eqs.(\ref{main 2}) and (\ref{Sbar}) the Fisher-Rao measure
$\gamma(\alpha)$ is
\begin{equation}
\gamma(\alpha)=\int d\theta\,\pi(\theta|\alpha)\left[  \frac{d}{d\alpha}%
\log\pi(\theta|\alpha)\right]  ^{2}=\frac{d^{2}\log\zeta(\alpha)}{d\alpha^{2}%
}\,, \label{gamma(alpha)}%
\end{equation}
and where $s(\alpha)$ is given by
\begin{equation}
s(\alpha)=-\int d\theta\,\pi(\theta|\alpha)\log\frac{\pi(\theta|\alpha
)}{g^{1/2}(\theta)}\,=\log\zeta(\alpha)-\alpha\frac{d\log\zeta(\alpha
)}{d\alpha}~. \label{s(alpha)}%
\end{equation}
This completes our derivation of the actual prior for $\theta$: the averaged
$\bar{\pi}(\theta)$ in eq.(\ref{main 3}) codifies information contained in the
likelihood function, plus the insight that for repeatable experiments,
information about the expected likelihood entropy, even if unavailable, is relevant.

\section{Example: a Gaussian model}

Consider data $y^{(N)}=\{y_{1},\ldots,y_{N}\}$ that are scattered around an
unknown value $\mu$,
\begin{equation}
y=\mu+\nu~
\end{equation}
with $\langle\nu\rangle=0$ and $\langle\nu^{2}\rangle=\sigma^{2}.$ The goal is
to estimate $\theta=(\mu,\sigma)$ on the basis of $y^{(N)}$ and the
information implicit in the data space $Y$, its measure $m(y)$ (discussed
below), and the Gaussian likelihood,
\begin{equation}
p(y|\mu,\sigma)=\frac{1}{\left(  2\pi\sigma^{2}\right)  ^{1/2}}\exp\left[
-\frac{(y-\mu)^{2}}{2\sigma^{2}}\right]  ~. \label{Gaussian likelihood}%
\end{equation}

We asserted earlier that knowing the measure $m(y)$ is part of knowing what
data has been collected. In many physical situations where the data happen to
be distributed according to eq.(\ref{Gaussian likelihood}) the underlying
space $Y$ is invariant under translations and we can assume
$m(y)=m=\operatorname{constant}$. Indeed, the Gaussian distribution can be
obtained by maximizing an entropy with an underlying constant measure and
constraints on the relevant information the mean $\mu$ and the variance
$\sigma^{2}$.

From eqs.(\ref{Stheta}) and (\ref{Gaussian likelihood}) the entropy of the
likelihood is
\begin{equation}
S(\mu,\sigma)=\log\left[  \frac{\sigma}{\sigma_{0}}\right]  \quad
\text{where}\quad\sigma_{0}\overset{\operatorname*{def}}{=}\left(  \frac
{e}{2\pi}\right)  ^{1/2}\frac{1}{m}~,
\end{equation}
and the corresponding Fisher-Rao measure, from eq.(\ref{Fisher metric}) is
\begin{equation}
g(\mu,\sigma)=\det\left\vert
\begin{array}
[c]{cc}%
1/\sigma^{2} & 0\\
0 & 2/\sigma^{2}%
\end{array}
\right\vert =\frac{2}{\sigma^{4}}~.
\end{equation}

Note that both $S(\mu,\sigma)$ and $g(\mu,\sigma)$ are independent of $\mu$.
This means that if we were concerned with the simpler problem of estimating
$\mu$\ in a situation where $\sigma$ happens to be known, the Bayesian
estimate of $\mu$ using entropic priors coincides with the maximum likelihood estimate.

When $\sigma$ is unknown the $\alpha$-dependent entropic prior,
eq.(\ref{main 2}), is
\begin{equation}
\pi(\mu,\sigma|\alpha)=\frac{2^{1/2}}{\zeta(\alpha)}~\frac{\sigma^{\alpha-2}%
}{\sigma_{0}^{\alpha}}~. \label{pi(theta|alpha)G0}%
\end{equation}
Since $\pi(\mu,\sigma|\alpha)$ is improper in both $\mu$ and $\sigma$ we must
introduce high and low cutoffs for both $\mu$ and $\sigma$. The fact that
without cutoffs the model is not well defined is interpreted as a request for
additional relevant information, namely, the values of the cutoffs.

We write the range of $\mu$ as $\Delta\mu=\mu_{H}-\mu_{L}$ and introduce
dimensionless quantities $\varepsilon_{L}$ and $\varepsilon_{H}$; $\sigma$
extends from $\sigma_{L}=\sigma_{0}\varepsilon_{L}$ to $\sigma_{H}=\sigma
_{0}/\varepsilon_{H}$. Then $\zeta(\alpha)$ and $\pi(\mu,\sigma|\alpha)$ are
given by
\begin{equation}
\zeta(\alpha)=\frac{2^{1/2}\Delta\mu}{\sigma_{0}}\frac{\varepsilon
_{H}^{1-\alpha}-\varepsilon_{L}^{\alpha-1}}{\alpha-1}~.
\end{equation}
and
\begin{equation}
\pi(\mu,\sigma|\alpha)=\frac{1}{\Delta\mu\sigma_{0}}\frac{\alpha
-1}{\varepsilon_{H}^{1-\alpha}-\varepsilon_{L}^{\alpha-1}}~\left(
\frac{\sigma}{\sigma_{0}}\right)  ^{\alpha-2}~.
\label{pi(theta|alpha)Gaussian}%
\end{equation}
Note that $\pi(\mu,\sigma|\alpha=1)$ reduces to $d\sigma/\sigma$ which is the
Jeffreys prior usually introduced by imposing invariance under scale
transformations, $\sigma\rightarrow\lambda\sigma$.

Writing $\varepsilon\overset{\operatorname*{def}}{=}(\varepsilon
_{L}\varepsilon_{H})^{1/2}$, the prior for $\alpha$, is obtained from
eq.(\ref{gamma(alpha)}),
\begin{equation}
\gamma(\alpha)=\frac{1}{(\alpha-1)^{2}}-\left(  \frac{2\log\varepsilon
}{\varepsilon^{1-\alpha}-\varepsilon^{\alpha-1}}\right)  ^{2}
\label{gammaGaussian}%
\end{equation}
and from eqs.(\ref{main 3}) and (\ref{s(alpha)}),
\begin{equation}
\pi(\alpha)=\frac{\gamma^{1/2}(\alpha)}{z}\frac{\varepsilon^{1-\alpha
}-\varepsilon^{\alpha-1}}{\alpha-1}\exp\left[  \frac{1}{\alpha-1}+\alpha
\frac{\varepsilon^{1-\alpha}+\varepsilon^{\alpha-1}}{\varepsilon^{1-\alpha
}-\varepsilon^{\alpha-1}}\log\varepsilon\right]  ~, \label{pi(alpha)Gaussian}%
\end{equation}
where the normalization $z$ has been suitably redefined.

\begin{figure}
[ptb]
\begin{center}
\includegraphics[
height=2.5554in,
width=3.2446in
]%
{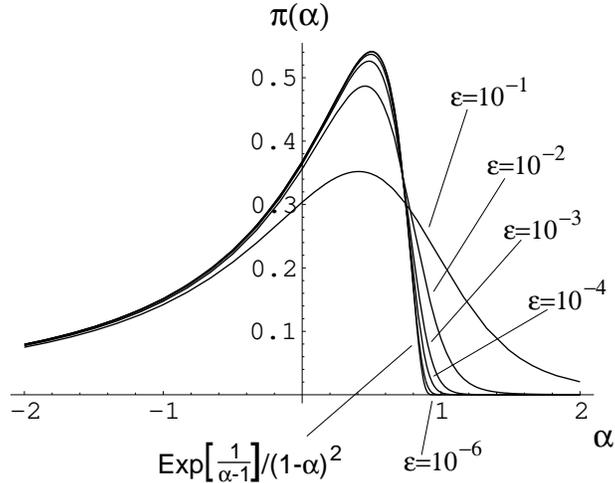}%
\caption{The prior $\pi(\alpha)$ for various values of the cutoff parameter
$\varepsilon$, as $\varepsilon\rightarrow0$. }%
\label{Fig. 1}%
\end{center}
\end{figure}

Eqs.(\ref{gammaGaussian}) and (\ref{pi(alpha)Gaussian}) simplify in the limit
$\varepsilon\rightarrow0$. Note that the same result is obtained irrespective
of the order in which we let $\varepsilon_{H}\rightarrow0$ and/or
$\varepsilon_{L}\rightarrow0$. The resulting $\gamma(\alpha)$ and $\pi
(\alpha)$ are%
\begin{equation}
\gamma(\alpha)=\frac{1}{(\alpha-1)^{2}}~,
\end{equation}
and
\begin{equation}
\pi(\alpha)=\left\{
\begin{array}
[c]{cc}%
\frac{1}{\left(  1-\alpha\right)  ^{2}}\exp\left[  {\frac{1}{\alpha-1}}\right]
& \text{for\quad}\alpha<1\\
0 & \text{for\quad}\alpha\geq1
\end{array}
\right.  \label{limit piGaussian}%
\end{equation}
where $\pi(\alpha)$ is normalized and is shown in Fig. \ref{Fig. 1}.%

$\pi(\alpha)$ reaches its maximum value at $\alpha=1/2$. Since $\pi
(\alpha)\sim\alpha^{-2}$ for $\alpha\rightarrow-\infty$ the expected value of
$\alpha$ and all higher moments diverge. This suggests that replacing the
unknown $\alpha$ in the prior $\pi(\theta|\alpha)$ by any given numerical
value $\hat{\alpha}$ is probably not a good approximation.\ 

\begin{figure}
[tb]
\begin{center}
\includegraphics[
height=2.6113in,
width=3.3829in
]%
{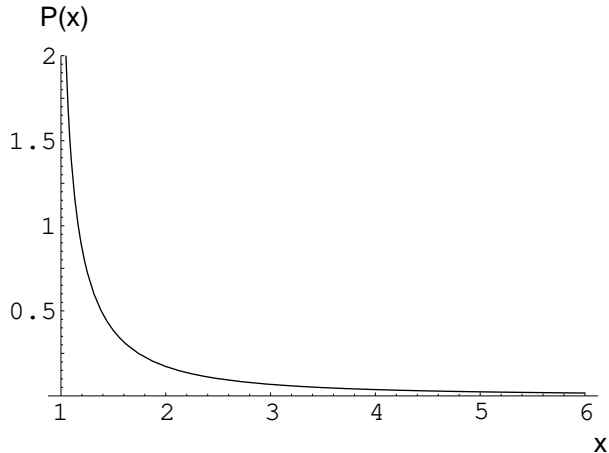}%
\caption{The effective $\bar{\pi}(\mu,\sigma)$ is shown as $P(x)=\frac{2}%
{x}K_{0}\left(  2\sqrt{\log(x)}\right)  $ where $x=\sigma/\sigma_{L}$. }%
\label{Fig. 2}%
\end{center}
\end{figure}

Since $\alpha$ is unknown the effective prior for $\theta=(\mu,\sigma)$ is
obtained marginalizing $\pi(\mu,\sigma,\alpha)=\pi(\mu,\sigma|\alpha
)\pi(\alpha)$ over $\alpha$, eq.(\ref{main 3}). Since $\pi(\alpha)=0$ for
$\alpha\geq1$ as $\varepsilon\rightarrow0$ we can safely take the limit
$\varepsilon_{H}\rightarrow0$ or $\sigma_{H}\rightarrow\infty$ while keeping
$\sigma_{L}$ fixed,
\begin{equation}
\pi(\mu,\sigma,\alpha)=\left\{
\begin{array}
[c]{cc}%
\frac{1}{\Delta\mu\sigma_{L}}\frac{\exp\left[  {\frac{1}{\alpha-1}}\right]
}{1-\alpha}\left(  \frac{\sigma}{\sigma_{L}}\right)  ^{\alpha-2} &
\text{for\quad}\alpha<1\\
0 & \text{for\quad}\alpha\geq1.
\end{array}
\right.  \label{pi(theta,alpha)G}%
\end{equation}
(However we cannot take $\sigma_{L}\rightarrow0$). The averaged prior for
$\mu$ and $\sigma$ is
\begin{equation}
\bar{\pi}(\mu,\sigma)=\frac{(\sigma_{L}/\sigma)^{2}}{\Delta\mu\sigma_{L}}%
\int_{-\infty}^{1}\frac{\exp\left[  {\frac{1}{\alpha-1}}\right]  }{1-\alpha
}\left(  \frac{\sigma}{\sigma_{L}}\right)  ^{\alpha}d\alpha=\frac{2}{\Delta
\mu\sigma}K_{0}\left(  2\sqrt{\log\frac{\sigma}{\sigma_{L}}}\right)  ~,
\label{main 5}%
\end{equation}
where $K_{0}$ is a modified Bessel function of the second kind. This is the
entropic prior for the Gaussian model. The function
\begin{equation}
P(x)=\frac{2}{x}K_{0}\left(  2\sqrt{\log x}\right)
\end{equation}
is shown in Fig. \ref{Fig. 2} as a function of $x=\sigma/\sigma_{L}$. The
singularity as $x\rightarrow1$ is integrable.%

\section{Final remarks}

Using the method of maximum relative entropy we have translated the
information contained in the known form of the likelihood into a prior
distribution. The argument follows closely the analogous application of the
ME\ method to statistical mechanics. For experiments that cannot be repeated
the resulting \textquotedblleft entropic prior\textquotedblright\ is formally
identical with the Einstein fluctuation formula. For repeatable experiments,
however, additional relevant information -- represented in terms of a Lagrange
multiplier $\alpha$ -- must be included in the analysis. The important case of
a Gaussian likelihood was treated in detail.

We have dealt with the simplest case where all we know about the quantities
$\theta$ is that they appear as parameters in the likelihood $p(y|\theta)$.
Our argument can, however, be generalized to situations where we know of
additional relevant information beyond what is contained in the likelihood.
Such information can be taken into account through additional constraints in
the maximization of the entropy $\sigma$.

To conclude we comment briefly on the entropic priors proposed by Skilling and
by Rodr\'{\i}guez. Skilling's prior, unlike ours, is not restricted to
probability distributions but is intended for generic \textquotedblleft
positive additive distributions\textquotedblright\ \cite{Skilling89}. Our
argument, which consists in maximizing the entropy $\sigma$ subject to a
constraint $p(y,\theta)=\pi(\theta)p(y|\theta)$, makes no sense for generic
positive additive distributions for which there is no available product rule.
Another important difference arises from the fact that Skilling's entropy is
not, in general, dimensionless and his hyper-parameter $\alpha$ is interpreted
some sort of cutoff carrying the appropriate corrective units. Difficulties
with Skilling's prior were identified in \cite{Skilling96}.

Rodr\'{\i}guez's approach is, like ours, derived from a maximum entropy
principle \cite{Rodriguez02}. One (minor) difference is his treatment of the
underlying measure $m(y)$. For us knowing $m(y)$ is part of knowing what data
has been collected; for him $m(y)$ is an initial guess and he suggests setting
$m(y)=p(y|\theta_{0})$ for some value $\theta_{0}$. The more important
difference, however, is that the number of observed data $N$ is left
unspecified. The space $\Theta\times Y^{N}$ over which distributions are
defined, and therefore the distributions themselves, also remain unspecified.
It is not clear what the maximization of an entropy over such unspecified
spaces could possibly mean but a hyper-parameter $\alpha$ is eventually
introduced and it is interpreted as a \textquotedblleft virtual number of
observations supporting the initial guess $\theta_{0}$.\textquotedblright\ A
different interpretation is given in \cite{Rodriguez03}. Since $\alpha$ is
treated on the same footing as the other parameters $\theta$,-
Rodr\'{\i}guez's approach requires an endless chain of hyper-parameters.

\noindent\textbf{Acknowledgments- }Many of our comments and arguments have
been inspired by Carlos C. Rodr\'{\i}guez, Volker Dose, and Rainer Fischer
through insightful questions and discussions which we gratefully acknowledge.
A. C. also acknowledges the hospitality of the Max-Planck-Institut f\"{u}r
Plasmaphysik during the two extended visits when most of this work was carried out.

\end{document}